\documentclass{aastex}          %% The manuscript based on AASTeX v5.x
\usepackage{spr-astr-addons}    %% mimicing ASTR journal style
\usepackage{xcolor}

\newcommand{\be}{\begin{equation}}
\newcommand{\ee}{\end{equation}}

\renewcommand\apjl{Astrophys. J. Lett.}%
\renewcommand\apjs{Astrophys. J. Suppl. Ser.}%
\renewcommand\ao{Appl. Opt.}%
\renewcommand\aap{Astron. Astrophys.}%
\newcommand{\maria}{\textcolor{black}}
\newcommand{\ikar}{\textcolor{black}}

\begin{document}
\graphicspath{{images/}}
%% Article title
%
\title{Local dark energy in the Sculptor Filament of galaxies}

%% Running heads
\shorttitle{Local dark energy in the Sculptor Filament}
\shortauthors{M.~V.~Pruzhinskaya et al.}

%% Author and Affilations
\author{M.~V.~Pruzhinskaya\altaffilmark{1}\altaffilmark{,*}} \and \author{A.~D.~Chernin\altaffilmark{1}}
\affil{*Corresponding author, pruzhinskaya@gmail.com}
\affil{Lomonosov Moscow State University, Sternberg Astronomical Institute, Universitetsky pr.~13, Moscow, 119234, Russia}
\and 
\author{I.~D.~Karachentsev\altaffilmark{2}}
\affil{Special Astrophysical Observatory RAS, Nizhnij Arhyz, 369167, Russia}
\email{pruzhinskaya@gmail.com} %% non-output

% Abstract
\begin{abstract}
Two dozens of different mass galaxies observed at distances less than 10 Mpc from the Local Group are organized in the elongated structure known as  the Sculptor Filament.  We use recent Hubble Space Telescope data on local galaxies  to study the dynamical structure and evolutionary trends of the filament. An N-body computer model, which reproduces its observed kinematics, is constructed under the assumption that the filament  is embedded in the universal dark energy background.  In the model, the motions of the filament members are controlled by their mutual gravity attraction force and the anti-gravity repulsion force produced by the local dark energy. It is found that the dark energy repulsion dominates the force field of the outer parts of the filament. Because of this, the filament expands and its expansion proceeds with acceleration.  The dark energy domination increases with cosmic time and introduces to the filament  the linear velocity--distance relation with the universal time-rate (``the Hubble constant") that depends asymptotically on the dark energy density only. 
\end{abstract}

% Keywords

\section{Introduction}
\label{s:intro}

The studies of the cosmic kinematics and dynamics started in the 1910s when Slipher discovered the recession motions of nearby galaxies. In 1922-24, Friedmann published his now famous  theory of the uniform universe. According to the theory, the universe as a whole  in not static, it is expanding, and its expansion motion follows the linear velocity-distance relation. The same linear relation known as the ``Hubble-Lema\^itre Law" was found  by Lema\^itre in 1927~\citep{1927ASSB...47...49L}, Robertson in 1928~\citep{Robertson1928}, and Hubble in 1929~\citep{1929PNAS...15..168H} in the observed motions of local (distances of $\sim$1--30~Mpc) galaxies. Is there any  fundamental interconnection among the local motions of galaxies  with the global expansion motion of the Universe as a whole?  The question has been opened  for decades. 

\maria{Friedmann's} global theory assumes that the matter distribution is uniform in the global universe. The real universe is uniform indeed, but this is the uniformity in average over the global space scales of $\sim$1000 Mpc and larger. This fact has been confirmed rather  recently, in 1990s, by observations with the Hubble Space Telescope and other modern astronomical instruments. As for the local space, the matter distribution is highly non-uniform there: the nearby galaxies are mostly collected in the building systems \maria{of} the Cosmic Web which are groups, clusters, superclusters (``Zeldovich \maria{pancakes}"), filaments of galaxies. Friedmann's theory is not applicable there; this theory cannot predict, explain or describe the local astronomical findings of the 1920s. The observed motions on both local and global space scales can be similar, only if they \maria{are} all essentially controlled by a common strong physical factor. This factor really exists: it is  dark energy. 

The solution to the long-standing problem~\citep{chernin2010} was suggested soon after the discovery of dark energy at the global cosmological distances~\citep{riess1998,perlmutter1999}. The solution is based on the following arguments:  the astronomers~\citep{riess1998,perlmutter1999} found dark energy indeed;  dark energy is described by  Friedmann's theory with Einstein's cosmological constant;  on the local space scales (1--30~Mpc), the anti-gravity of dark energy is as strong as the  gravity produced by the matter of galaxies there. The first two (``global") arguments are confirmed by all the development of cosmology for the last two decades. As for the third (``local") argument, it includes dark energy to the local dynamics of galaxies  in a natural way because Einstein's cosmological constant is constant  and so dark energy has the same density on  both global and local scales\footnote{The tension between the global and local $H_0$ measurements~\citep{Planck2018,2019ApJ...876...85R} can be due to some unknown systematics and not necessary includes the  additional physics beyond  the current  standard  cosmological  model and discussed in~\cite{2019ApJ...882...34F}.}. This argument is supported by a set of computer models developed for really existing systems of galaxies embedded in the dark energy omnipresent background. These systems are as following: the Local Group (LG) of galaxies together with the expansion outflow of dwarf galaxies around it, several systems similar to the Local Group, two clusters of galaxies (Coma and Virgo) with their expansion outflows, the Local  Supercluster of galaxies, or ``Zeldovich pancake" (see \citealt{chernin2001,chernin2008,2013JETPL..98..353C,chernin2015}, and references therein). 

The set of the models is not complete. Indeed, the three types of galaxy systems are known as the basic building blocks of the observed Cosmic Web. They are the three-dimensional (3D) groups and clusters, two-dimensional (2D) walls like the Local Supercluster and one-dimensional (1D) filaments. In our set of the local models, 3D and 2D systems are represented and there is no 1D systems which are very typical elements of the large-scale structure~\citep{2014MNRAS.438.3465T}. To fill this lack, a $\Lambda $N-body model is suggested in the present work. This is the model of the nearby Sculptor Filament of galaxies, \maria{a loose filament of galaxies located in the immediate proximity of the Local Group (see Sec.~\ref{Scu})}.  The filament is an expanding system, and it will be shown \maria{below} that anti-gravity of dark energy plays an important role in its dynamics. 
Because of this, the expansion of the system proceeds with acceleration approaching with time the linear velocity-distance relation. 

The basic astronomical data for the theory are provided by the Hubble Space Telescope (HST) observations~\citep{2003A&A...404...93K,2013AJ....145..101K}.  
 
Note that more than half a century ago,~\cite{1959ApJ...130..718D} described the Sculptor Filament as an ``expanding association of galaxies" which ``has a total positive energy ``being" of the type predicted by Ambartsumian". Now one may see that the  physical nature of the filament expansion  is  rather due to omnipresent dark energy (which is hardly less mysterious than the ``superdense D-bodies"\footnote{\maria{One of the concepts of the formation of stars and galaxies existed in~1960--1970s. In particular, according to this concept clusters of galaxies arrise from a number of explosions of superdense protostellar objects, D-bodies~(\citealt{1958IzArm..11....9A,1958RvMP...30..944A,1962PrCmg...8....2A}; see also~\citealt{1971wssp.conf...37C}).}},  but nevertheless really existing in Nature).

\section{Dark energy on local scales}

According to the concept suggested by~\cite{1966JETP...22..378G}, dark energy may be treated topically as a vacuum-like continuous medium. This medium cannot serve as a reference frame, since it is comoving to any matter motion --- similarly to trivial emptiness. Its density is perfectly uniform, the same in any point of the space and any moment of time. The dark energy density is given by Einstein's cosmological constant:  
% 1 
\be \rho_{\Lambda} = {\Lambda}/(8 \pi G),\ee 
where $G$ is the Newtonian gravitational constant; the speed of light $c = 1$ here. Dark energy density $\rho_{\Lambda}$ is positive and its currently adopted value $\rho_{\Lambda} \simeq 0.7 \times 10^{-29}$ g/cm$^{3}$~\citep{riess1998,perlmutter1999,2018ApJ...859..101S}. The dark energy equation of state is 
% 2
\be p_{\Lambda} = - \rho_{\Lambda}.\ee

\noindent Here $p_{\Lambda}$ is the dark energy pressure which is negative, while  its  density is positive. 

General Relativity (GR) indicates that the ``effective gravitating density" of any medium is determined by both density and pressure of the medium:
% 3
\be \rho_{eff} = \rho + 3 p.\ee

The effective density of dark energy, 
% 4
\be \rho_{\Lambda} + 3 p_{\Lambda} = - 2 \rho_{\Lambda} < 0, \ee 
is negative, and it is because of this sign ``minus" that dark energy produces not attraction, but repulsion, or anti-gravity.

A simple example of the GR exact solution with non-zero dark energy is the 
Schwarzschild-de Sitter spacetime which gives the metric outside a spherical matter mass $M$ embedded in the dark energy of the constant density $\rho_{\Lambda}$:
% 5
\be ds^2 = Y dt^2 - R^2 d\Omega^2 - Y^{-1} dR^2. \label{eq:5} \ee

\noindent Here $R$ is the distance from the center of the mass,
$d\Omega^2 = \sin^2 \theta d\phi^2 + d\theta^2$ and
% 6
\be Y(R) = 1 - 2\frac{GM}{R} - \frac{8\pi}{3}\rho_{\Lambda} R^2.
\label{eq:6} \ee

The metric of equations~(\ref{eq:5},~\ref{eq:6}) is static, contrary to Friedmann's time-dependent spacetime. As it is seen from  equation~(\ref{eq:6}),  the gravity of the mass $M$ becomes negligible in
comparison with the dark energy anti-gravity, at infinitely large distances. In this space limit, the metric tends to de Sitter's metric which is determined
by dark energy only:
% 7
\be Y(R) \simeq 1 - \frac{8\pi}{3}\rho_{\Lambda} R^2, \;\;\; R
\rightarrow \infty. \label{eq:7} \ee

\noindent It is remarkable that Friedmann's cosmological metric with dark energy 
has the same de Sitter static asymptotic at infinitely large time in standard cosmology.

In the Newtonian terms, equations~(\ref{eq:5},~\ref{eq:6}) describe the gravity-anti-gravity force field, where Newtonian gravity is produced by the mass $M$ and Einstein's anti-gravity is produced by dark energy. If the force field is weak, so that  deviations from the Galilean metric are small, the metric of equation~(\ref{eq:7}) may be reduced to the Newtonian description in terms of the gravity-antigravity potential $U$:
% 8
\be Y^{1/2} \simeq 1 + U, \;\;\; U(R) = -\frac{GM}{R} - \frac{4\pi
G}{3} \rho_{\Lambda} R^2. \label{eq:8} \ee

In this approximation, the force (per unit mass) comes from equation~(\ref{eq:8}):
% 9
\be F(R) = - \frac{d U}{d R} = - \frac{GM}{R^2} + \frac{8\pi G}{3}
\rho_{\Lambda} R. \label{eq:9} \ee

\noindent In the rhs here the sum of the Newtonian force of gravity and Einstein's force of anti-gravity are given for unit mass, i.e. acceleration. 

It is seen from equation~(\ref{eq:9}) that gravity may dominate at small distances from the mass $M$, while anti-gravity may be stronger than gravity at large distances. Gravity and antigravity balance at the distance
% 10
\be R = R_{\Lambda} = \left(\frac{M}{\frac{8\pi}{3}
\rho_{\Lambda}}\right)^{1/3}.  \label{eq:10} \ee

The radius $R_{\Lambda}$  is the radius of the ``zero-gravity sphere"~\citep{chernin2001,2016BaltA..25..296C}. It appears here as the local spatial counterpart of the ``zero-gravity moment" in the global expansion of the Universe which occurred at the cosmic redshift 0.7 about 7~Gyr ago.

The Newtonian approximation of equations~(\ref{eq:7}--\ref{eq:10})  is appropriate for local systems, since the velocities of the expansion flows are very small compared to the speed of light, and the spatial differences in the gravity-antigravity potential are very small compared to the speed of light squared there. This approximation was used in the 3D and 2D models of the Cosmic Web systems (see \citealt{chernin2001,chernin2008, chernin2015}, and references therein). It is also used below for the 1D model of the Sculptor Filament of galaxies.  

\section{Sculptor Filament: 1D local expansion flow} 
\label{Scu}

\begin{figure}
	\includegraphics[trim={2.5cm 0.5cm 0.5cm 1.5cm},clip,width=1\linewidth]{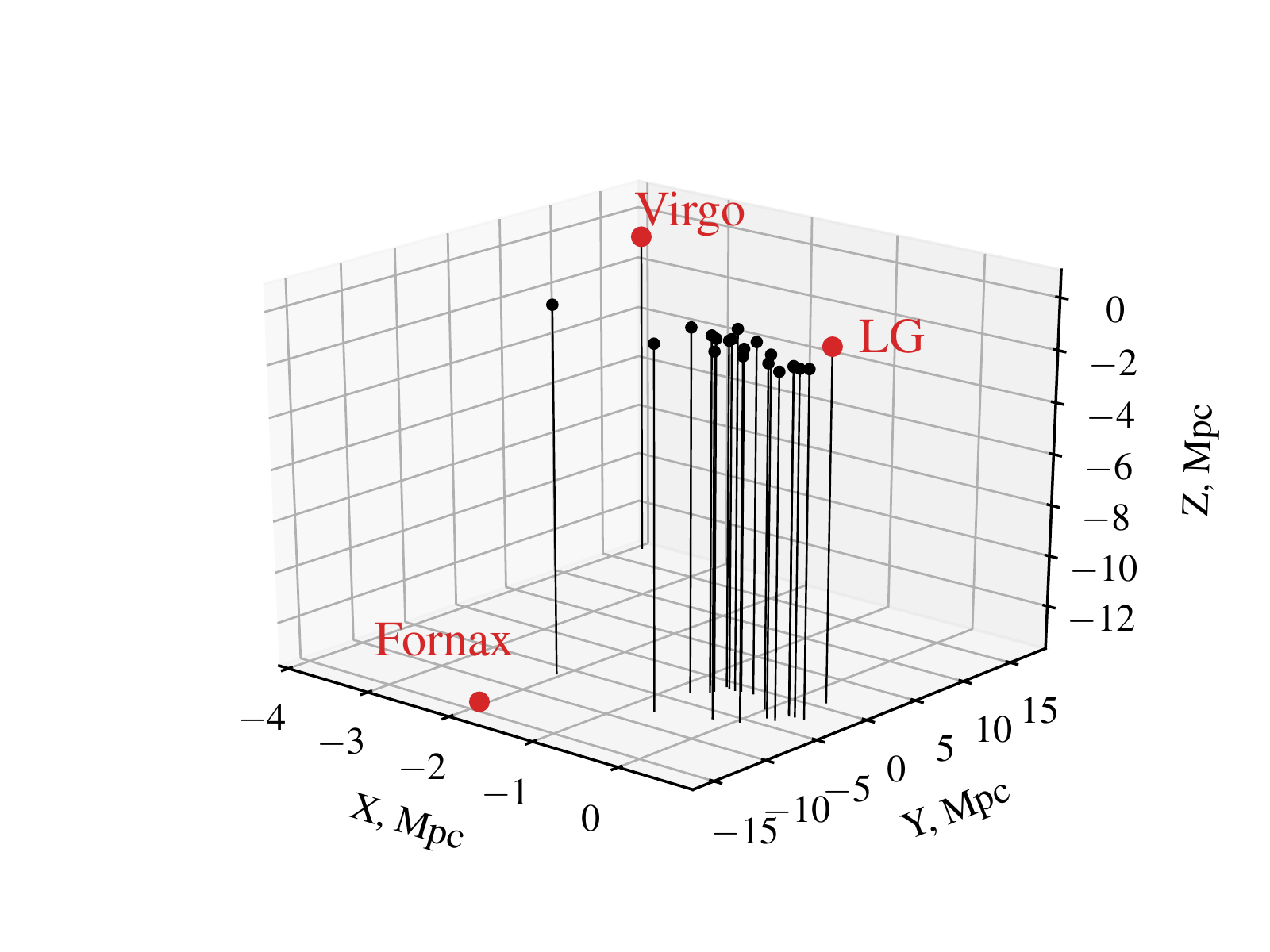}
    \caption{\maria{Positions of the Sculptor Filament and the nearby Virgo and Fornax clusters in Cartesian Supergalactic coordinates.}}
    \label{fig:1}
\end{figure}

\maria{The closest to the Local Group the association of the bright galaxies (NGC0055, NGC0247, NGC0253, NGC0300, and NGC7793) including a number of dwarf galaxies is known as the Sculptor group. However, it turns out that the Sculptor group does not constitute a single, compact, gravitationally-bound group, but rather to be a loose filament of galaxies of $1\times6$ in size extended along a line of sight over $\sim5$~Mpc~\citep{1985AJ.....90.1012A,1998AJ....116.2873J,2003A&A...404...93K,2004AJ....127.2031K}. Following~\cite{2017MNRAS.472.4832W} we use the term ``Sculptor Filament" for it throughout this paper. In fact, the Local Group forms a part of this filament~\citep{2003A&A...404...93K}. Due to our location within the Sculptor Filament, we see the sub-groups of the filament lined up along the line of sight like a ``cigar".  It is needed to note that the considered filament is a very small structure in comparison with the huge filaments such as the Perseus-Pisces filament~\citep{1993AJ....105.1251W} or the Norma-Pavo-Indus filament~\citep{2013AJ....146...69C}. The two nearest clusters of galaxies to the Sculptor Filament are the Virgo and Fornax clusters (see~Fig.~\ref{fig:1}).}

It was also recognized that the near and the far parts of the ``cigar" are 
not gravitationally bound of each other and follow linear (Hubble-Lema\^itre) 
velocity-distance relation.   

List with the main members of the Sculptor Filament is presented in~Table~\ref{tab:scu_table}. \maria{The object names correspond to the ones given in the Database on the Local Volume
Galaxies\footnote{\ikar{http://www.sao.ru/lv/lvgdb}}~\citep{2012AstBu..67..115K}.} The coordinates in the table are equatorial \ikar{(J 2000.0)}. The mass $M_\star$ is the total stellar mass \maria{obtained from the galaxy luminosity in $K$-band, $L_K$, assuming a ratio of $M_\star/L_K\simeq 1~M_\odot/L_\odot$~\citep{2003ApJS..149..289B}}. The radial velocity $V_{LG}$  and the distance $R$ are also given  in~Table~\ref{tab:scu_table}. The last column contains the method used to measure the galaxy distance: the tip of red giant branch (TRGB), the Tully-Fisher relation (TF), \maria{the numerical action method (NAM). The data are available in the Database on the Local Volume Galaxies and also published in~\cite{2013AJ....145..101K,2014AJ....147...13K,2017AJ....153....6K,2017MNRAS.472.4832W}.}

 \begin{table}
 \caption{Galaxies in the Sculptor Filament and around it (N=31)} %% no full stop at the end of caption
 \label{tab:scu_table}
 \scalebox{0.69}{\begin{tabular}{lccccr}
 \tableline  %% rule at top
% \tablenotemark{a} 
		 Name & Coord.   &  log$_{10}$ $M_\star$      & $V_{LG}$ & $R$    & meth\\
		    	   & \ikar{RA}, Dec &  M$_\odot$ & km s$^{-1}$    & Mpc& \\
\hline
ESO409-015 &     00 05 31.8 -28 05 53  &8.10   &769 &   8.71 &TRGB\\
ESO349-031 &    00 08 13.3 -34 34 42  &7.12   &230  & 3.21 &TRGB\\
NGC0024      &   00 09 56.4 -24 57 48  &9.48   &606   &7.31 &TRGB\\
NGC0045       &  00 14 03.9 -23 10 56  &9.33   &528   &6.64 &TRGB\\
NGC0055       &  00 15 08.5 -39 13 13  &9.48   &111   &2.11 &TRGB\\
NGC0059        & 00 15 25.1 -21 26 38  &8.66   &431   &4.90 &TRGB\\
ESO410-005     & 00 15 31.4 -32 10 48 & 6.89   & 53   &1.93 &TRGB\\
LV J0015-3825   &00 15 38.3 -38 25 41  &6.68&  592  &7.31&NAM\\
Sc22           & 00 23 51.7 -24 42 18  &7.15    &-    &4.29 &TRGB\\
ESO294-010 &     00 26 33.3 -41 51 20 & 6.30    &71&   2.03 &TRGB\\
ESO473-024   &   00 31 22.5 -22 45 57  &7.72  & 584 &  9.90 &TF\\
ScuSR         &  00 33 51.8 -27 50 24  &6.36    &-    &4.00 &mem\\
DDO226         & 00 43 03.8 -22 15 01  &7.71   &409 &  4.92& TRGB\\
NGC0247        & 00 47 08.3 -20 45 36  &9.50   &216  & 3.72 &TRGB\\
NGC0253         &00 47 34.3 -25 17 32 &10.98   &276  & 3.70 &TRGB\\
Scl-MM-Dw1     & 00 47 34.9 -26 23 20  &6.77   & -    &3.94 &TRGB\\
KDG002          &00 49 21.1 -18 04 28  &6.85   &290 & 3.56 &TRGB\\
DDO006         & 00 49 49.3 -21 00 58  &7.08   &344   &3.44 &TRGB\\
Scl-MM-Dw2     & 00 50 17.1 -24 44 59  &7.36    &-    &3.12 &TRGB\\
ESO540-032      &00 50 24.6 -19 54 25  &6.83  & 285 & 3.63& TRGB\\
NGC0300         &00 54 53.5 -37 40 57  &9.41   &116   &2.09 &TRGB\\
LV J0055-2310  & 00 55 01.0 -23 10 09  &6.17   &288&  3.70 &mem\\
UGCA438         &23 26 27.5 -32 23 26  &7.59    &99  &2.22 &TRGB\\
ESO347-017     & 23 26 56.1 -37 20 49 & 8.17&  701  &7.60 &TF\\
IC5332          &23 34 27.5 -36 06 06  &9.62   &716  &7.80& mem\\
LV J2335-3713&   23 35 04.1 -37 13 14 & 7.38  & 623&   7.59& TRGB\\
NGC7713         &23 36 15.0 -37 56 20  &9.43   &696 &7.80 &TF\\
PGC680341      & 23 41 47.5 -33 08 41 & 7.15  & 486  & 6.08&NAM\\
UGCA442         &23 43 46.0 -31 57 33  &8.03   &300  & 4.37& TRGB\\
NGC7793         &23 57 49.4 -32 35 24  &9.70   &250  &3.63 &TRGB\\
PGC704814       &23 58 40.7 -31 28 03  &6.90 &  299  & 3.60 &mem\\
 \tableline %% rule at bottom
 \end{tabular}}
 \end{table}

\section{$\Lambda$N-body model}
\subsection{Equations of motion}
Following the same logic as in~\cite{chernin2015}, in our $\Lambda$N-body model, the galaxies in Sculptor Filament \maria{(with and without the Virgo and Fornax clusters)} are treated as a non-relativistic isolated conservative system of point-like masses interacting with each other via Newton's mutual gravity and undergoing Einstein's antigravity produced by the dark energy background. The equations of motion for the system can be written in the following form:
\begin{equation}
\frac{d{}^2x_i}{dt^2} = G \sum_{j=1, j\neq i}^{N} m_j\frac{x_j - x_i}{r^3_{ij}} + H^2_\Lambda x_i
\label{eq:11}
\end{equation}

\begin{equation}
\frac{d{}^2y_i}{dt^2} = G \sum_{j=1, j\neq i}^{N} m_j\frac{y_j - y_i}{r^3_{ij}} + H^2_\Lambda y_i
\label{eq:12}
\end{equation}

 \begin{equation}
\frac{d{}^2z_i}{dt^2} = G \sum_{j=1, j\neq i}^{N} m_j\frac{z_j - z_i}{r^3_{ij}} + H^2_\Lambda z_i
\label{eq:13}
\end{equation}

\begin{equation}
r_{ij} = \sqrt{(x_j - x_i)^2+(y_j - y_i)^2+(z_j - z_i)^2}
\label{eq:14}
\end{equation}

\begin{figure*}[h]
\begin{minipage}[h]{0.49\linewidth}
\center{\includegraphics[width=1\linewidth]{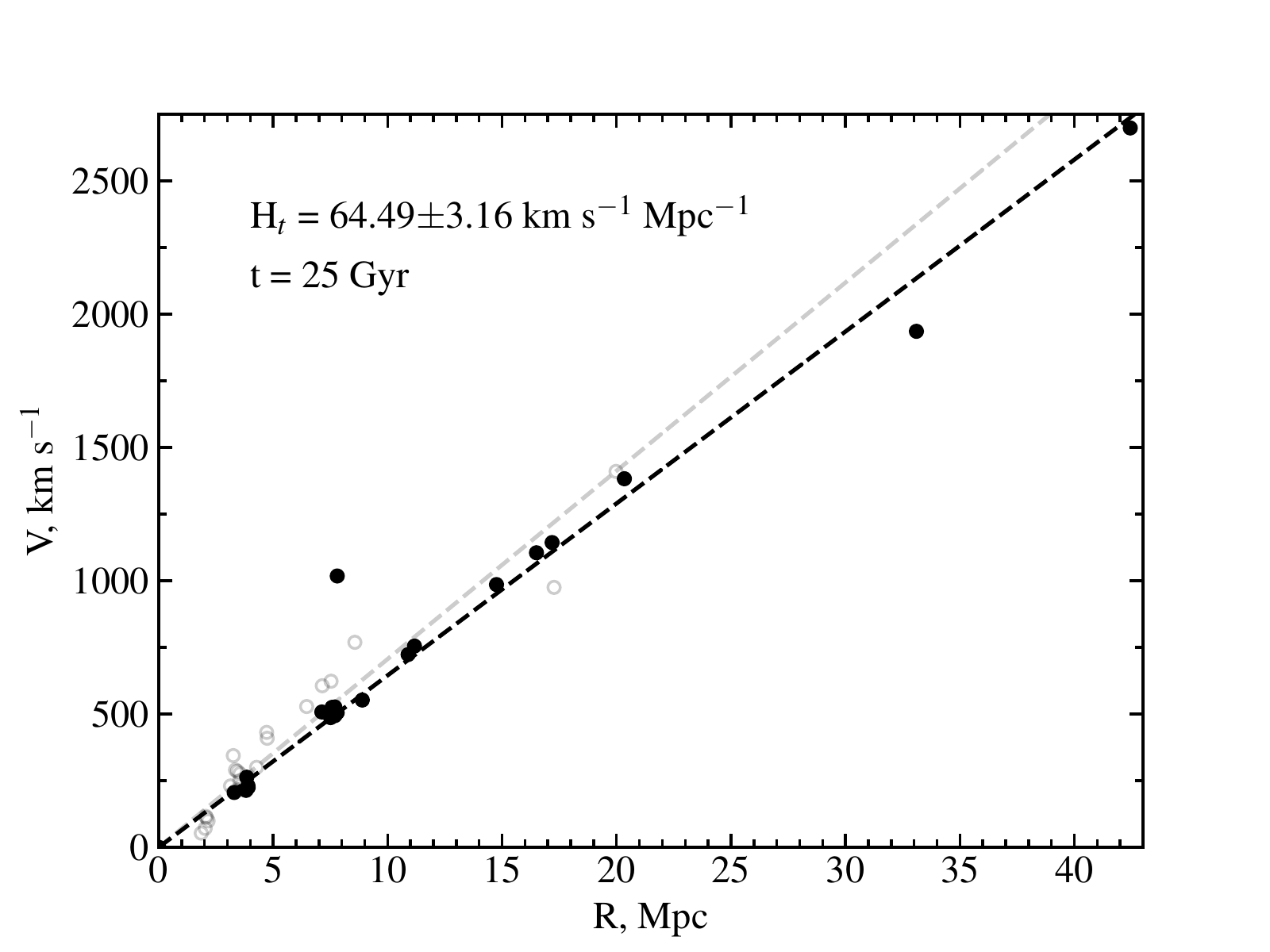} \\ (a)}
\end{minipage}
\hfill
\begin{minipage}[h]{0.49\linewidth}
\center{\includegraphics[width=1\linewidth]{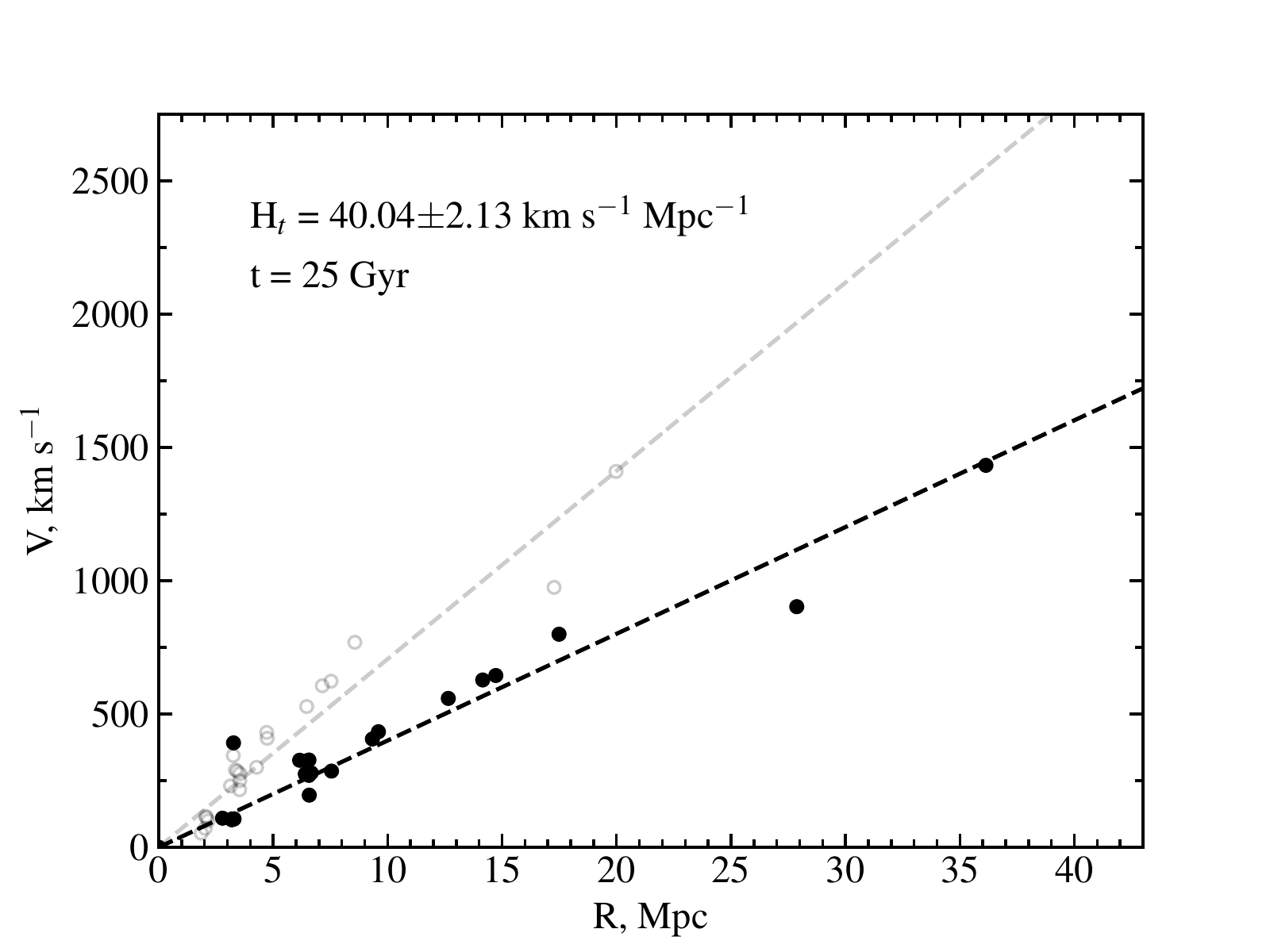} \\ (b)}
\end{minipage}
\caption{Velocity-distance diagrams for the Sculptor Filament and \maria{the Virgo and Fornax clusters}  with (a) and without (b) dark energy term in the equations of motions for $t =  25$~Gyr. The dashed line is the fit by linear relation $V = H_t R$. The Hubble diagram based on the velocity and distance measurements in the current epoch ($t = 13.8$~Gyr) is shown by light-grey colour in both plots.}
\label{fig:HD}
\end{figure*}

To work with more homogeneous set of the data for our modelling we used only the members of the Sculptor Filament with  known velocities and the TRGB distances, i.e. 19 galaxies~(see Table~\ref{tab:scu_table}). We performed the calculations in the supergalactic ($X, Y, Z$) coordinate system. The Local Group with the Milky Way and the Andromeda galaxies is considered as one point-like source with the total mass $M=3.11\times10^{12}$~$\rm M_\odot$. The total mass of other galaxies is estimated from the stellar mass with use of the relation  ${\rm log_{10}}(M_{tot}/M_\star) = {\rm log_{10}}(32) - 0.50\times {\rm log_{10}}(M_\star/10^{10})$, when ${\rm log_{10}}(M_\star) < 10$~\citep{2017ApJ...843...16K}. The total mass for NGC0253 is adopted to  $1.51\times10^{12}$~$M_\odot$~\citep{chernin2015}. \maria{Our main model includes the Virgo and Fornax clusters since they are located at the opposite sides of the filament and can potentially affect the flow. We also performed the simulations without these clusters. Their radial velocities are taken equal to 975 and 1410~km~s$^{-1}$, the total masses --- to $8\times10^{14}$ and $1\times10^{14}$~$M_\odot$, and the distances --- to 17 and 20 $\rm{Mpc}$, respectively~\citep{2010MNRAS.405.1075K}.}

\maria{We do not consider the origin and early evolution of the filament in the past. We use the observed velocities and distances of the galaxies at the present moment of cosmic time $t = t_0 = 13.8$~Gyr~\citep{Planck2016} as initial conditions in our model. They are listed in Table~\ref{tab:scu_table}.} Our model covers the time interval from $t_0$ to $t = 25$ Gyr.

The universal time-rate $H_\Lambda= \left(\frac{8\pi}{3}G\rho_{\Lambda}\right)^{1/2}$ is adopted to be 61 km~s$^{-1}$~Mpc$^{-1}$.  This time-rate is constant and depends on the dark energy density only (see~\citealt{chernin2015} for details).

It should be noted that the tangential (transverse) velocities of the galaxies are unknown and we allow these to be zero in the initial conditions for the model\footnote{\maria{The Gaia end-of-mission proper motions will be able to significantly detect the mass distribution of large-scale structure on length scales $<25$~Mpc which in turn will help us to determine the transverse peculiar velocities of galaxies and to improve our model in further studies~(\citealt{2018ApJ...868...69T}, see also~\citealt{2017ApJ...850..207S}).}}. Non-zero transverse velocities might change the trajectories of the flow in some way. However, they could hardly alternate the main trends of the system evolution and especially the asymptotic state of the flow because the tangential velocities vanish with the growth of the distances.

\begin{figure*}[h]
\begin{minipage}[h]{0.49\linewidth}
\center{\includegraphics[width=1\linewidth]{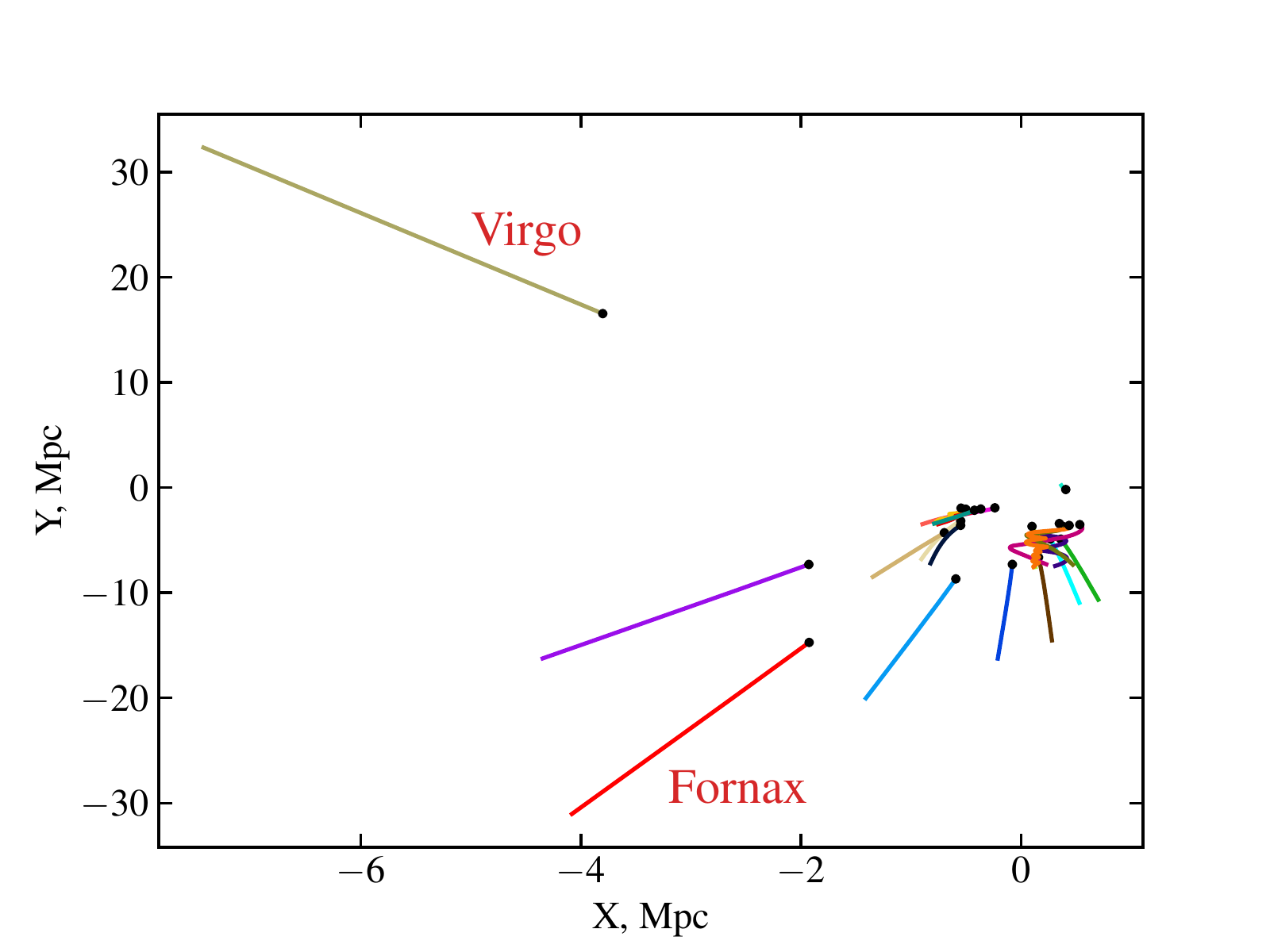} \\ (a)}
\end{minipage}
\hfill
\begin{minipage}[h]{0.49\linewidth}
\center{\includegraphics[width=1\linewidth]{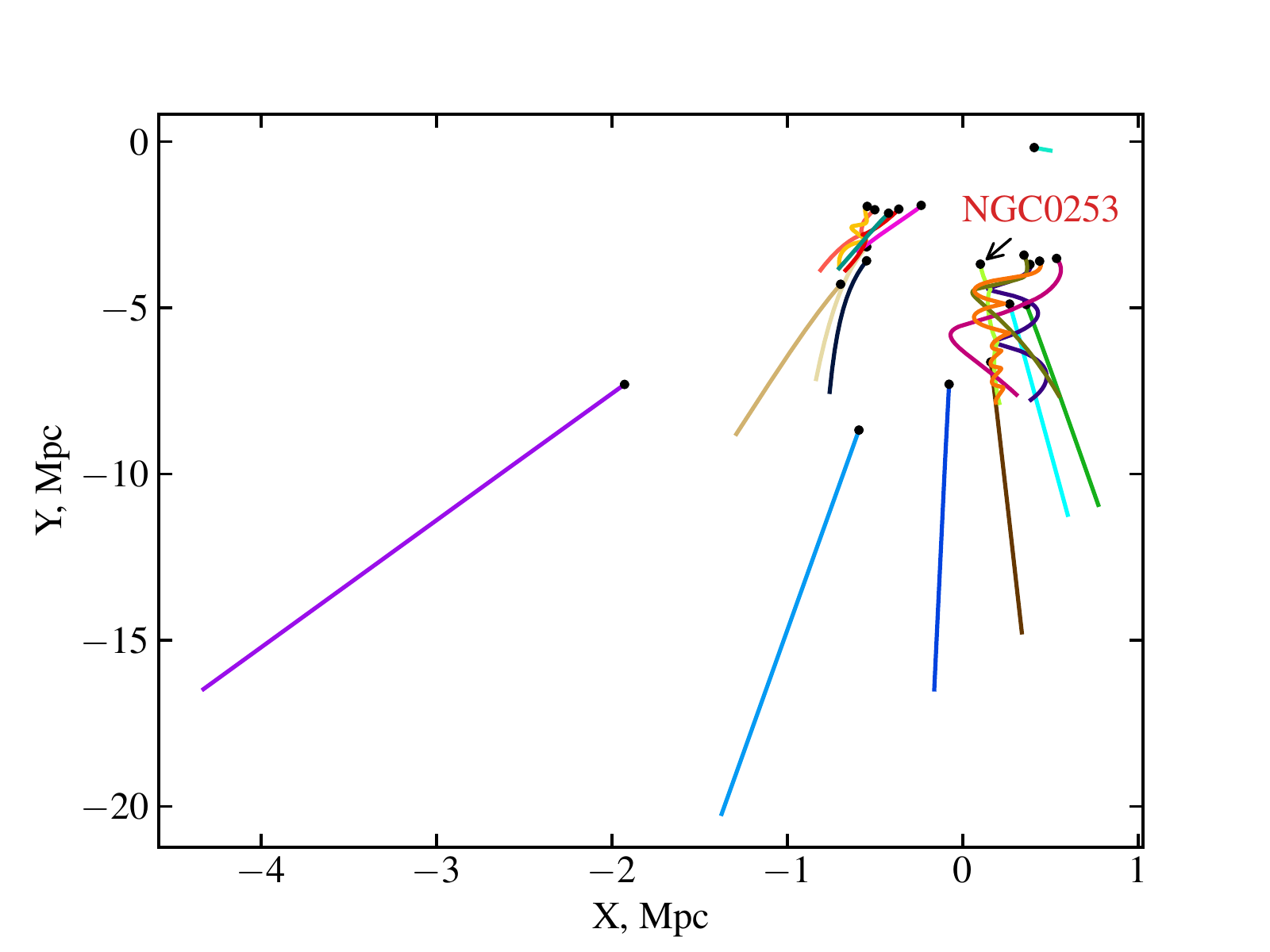} \\ (b)}
\end{minipage}
\caption{\maria{Spatial trajectories of the Sculptor Filament in the projection to the $XY$ supergalactic plane with (a) and without (b) the Virgo and Fornax clusters}. Black dots denote the present-day values.}
\label{fig:2}
\end{figure*}

\subsection{System dynamics and evolution}

In Fig.~\ref{fig:HD} we compare two Hubble diagrams, with and without dark energy term in equations~(\ref{eq:11}--\ref{eq:14}), at the moment $t=25$~Gyr. \maria{For this figure, the radial distances and velocities are re-calculated to the reference frame of the Local Group barycentre.} The initial positions of galaxies are shown by light-grey colour in both plots, the final positions are coloured in black. If we take into account the dark energy term, in the end of calculations the trajectories converge to the straight line with $H_t = 64.49\pm3.16$~km~s$^{-1}$~Mpc$^{-1}$. This value is not very far from the present value of the time-dependent cosmological time-rate $H_0=67.8\pm0.9$~km~s$^{-1}$~Mpc$^{-1}$ found by Planck~\citep{Planck2016}. On the contrary, without dark energy $H_t \simeq 40$~km~s$^{-1}$~Mpc$^{-1}$ which  does not match the current measurements. \maria{If we do not take into account the Virgo and Fornax clusters, the $H_t$ parameter equals to  $\sim67$~km~s$^{-1}$~Mpc$^{-1}$ (with dark energy term) and $\sim42$~km~s$^{-1}$~Mpc$^{-1}$ (without dark energy term).}

It can be also noted that the length of the phase trajectories is shorter for the bodies with lower initial velocities and longer for the high-velocity galaxies. The initially slowest galaxy --- ESO410 --- increases its distance from 1.9 to \maria{3.3 Mpc (1.74 times)} during this time interval. The fastest galaxy --- ESO409 ---  increases its distance from 8.7 to \maria{20.3~Mpc (2.33 times)}.

As we can see from Fig.~\ref{fig:HD} the radial velocity dispersion for the outer part of the filament at the moment $t=25$~Gyr is smaller than at the present moment which means that the flow there becomes increasingly regular and cold. However, the inner part of the filament keeps its dispersion and does not follow the global expansion caused by dark energy.

The $\Lambda N$-body model gives the spatial trajectories of galaxies in the Sculptor Filament (\maria{see their projections to the $XY$} supergalactic plane \maria{for our main model and the model that does not include the Virgo and Fornax clusters} in Fig.~\ref{fig:2}). In the figure we denote NGC0253, which is the most massive object in the filament after the MW-Andromeda system. In general, the distances of the bodies are increasing during the whole time of computation, and the mutual distances between them are also increasing. However, around NGC0253 the trajectories do not look like straight lines and affected by its gravity. Apparently, in this region the system experiences the multiple mutual passages and the point-like approximation does not work here. This explains the dispersion we observe for the inner part of the filament in Fig.~\ref{fig:HD}.

\begin{figure*}[h]
\begin{minipage}[h]{0.49\linewidth}
\center{\includegraphics[width=1\linewidth]{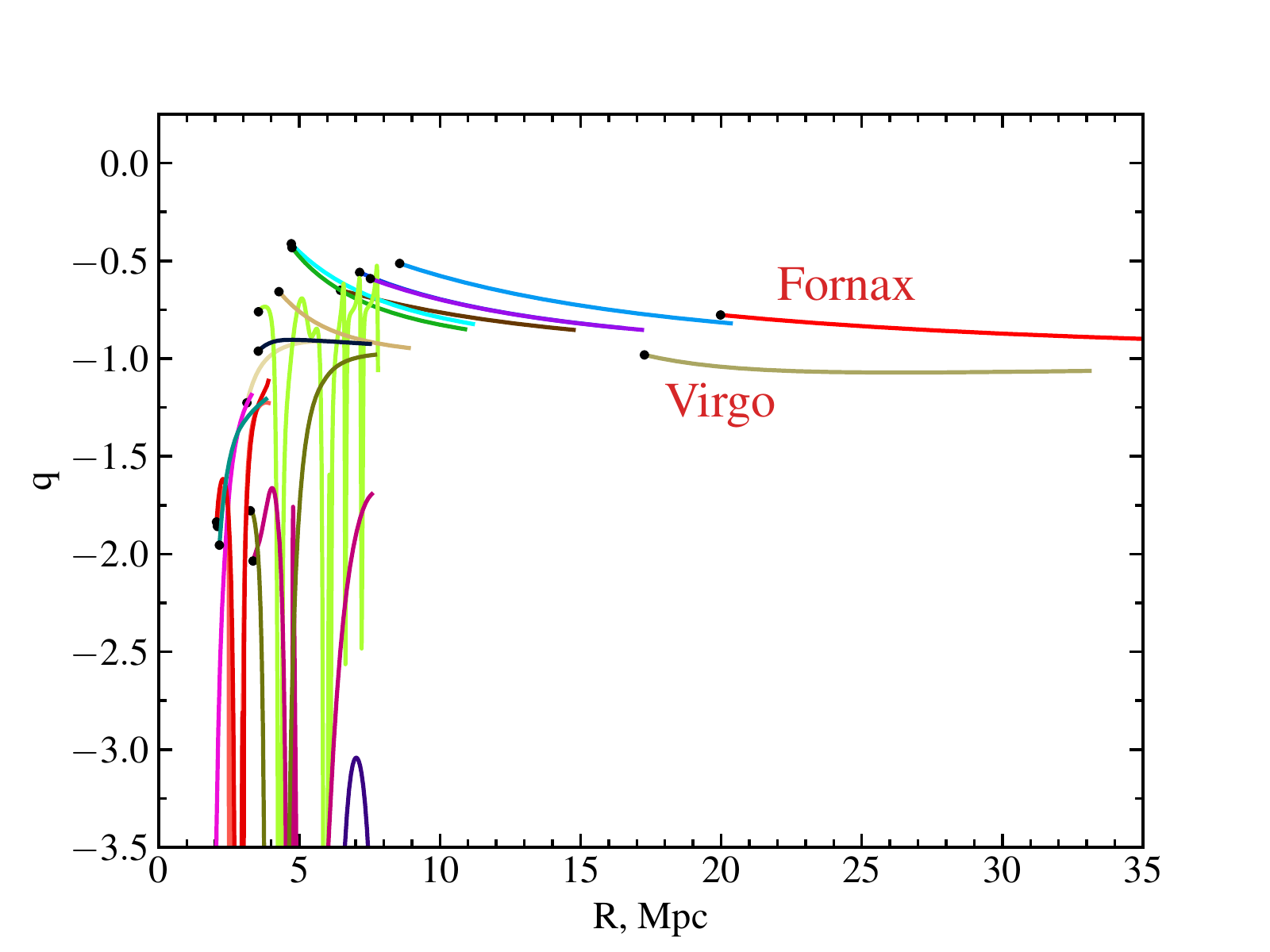} \\ (a)}
\end{minipage}
\hfill
\begin{minipage}[h]{0.49\linewidth}
\center{\includegraphics[width=1\linewidth]{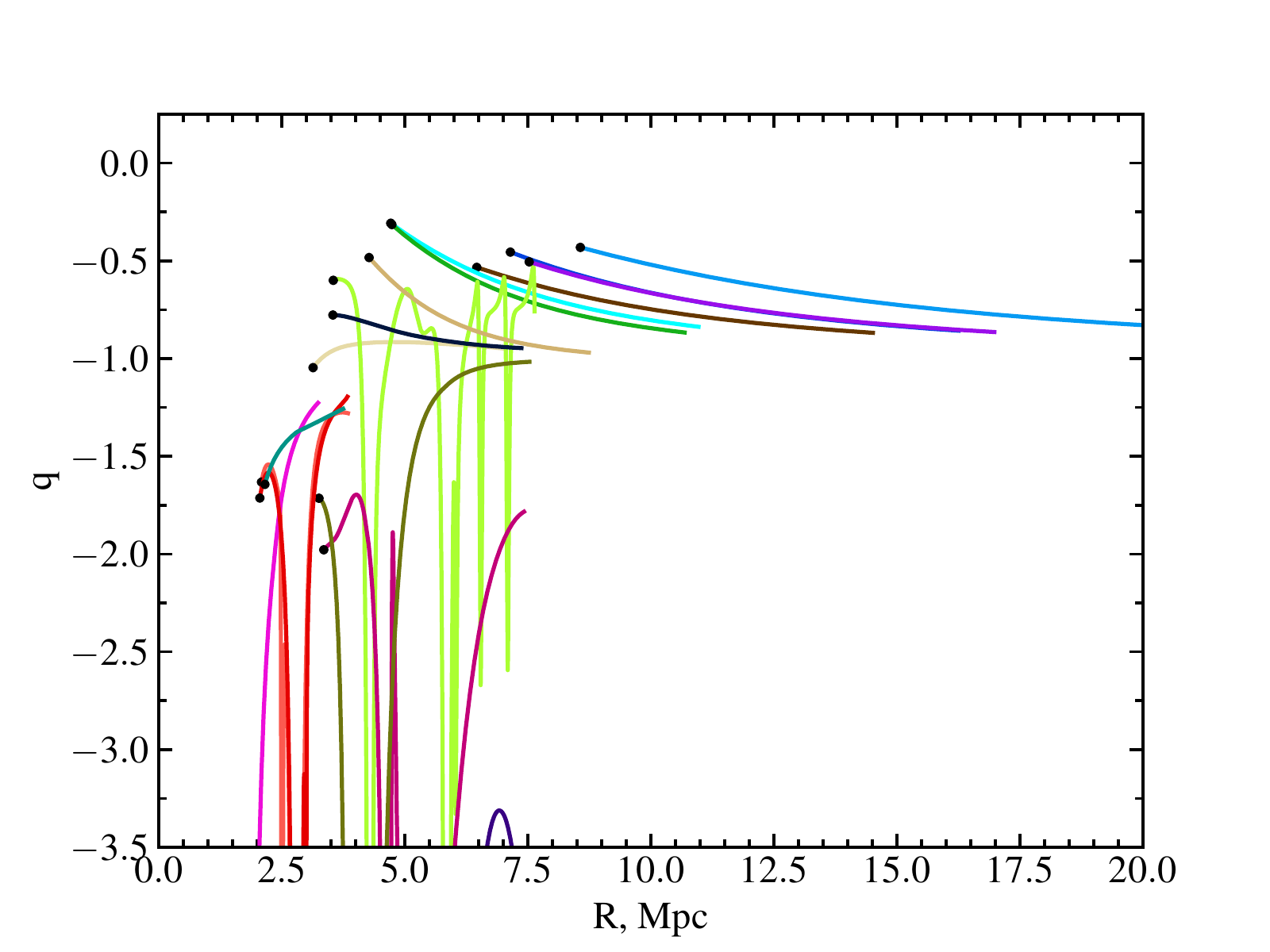} \\ (b)}
\end{minipage}
\caption{Deceleration parameter $q(R)$ as a function of the radial distance of the Sculptor Filament \maria{with (a) and without (b) the Virgo and Fornax clusters}. Black dots denote the present-day values.}
\label{fig:4}
\end{figure*}

\subsection{Acceleration}
The dimensionless deceleration parameter is defined as

\begin{equation}
q(t) = - \frac{\ddot{a}a}{\dot{a}^2},
\end{equation}
where a(t) is the cosmological scale factor. The cosmological deceleration parameter is positive for gravity-dominated Universe and negative for antigravity--dominated Universe tending to $q(t)_\infty=-1$ in the limit of infinite cosmological expansion. At the present epoch, the cosmological deceleration parameter $q_0 \simeq -0.54$~\citep{Planck2016}.

Let us introduce the deceleration parameter $q(R)$ as a function of the radial distance $R$ for the galaxies in the Sculptor Filament~\citep{2013A&AT...28..177C}:

\begin{equation}
q(t) = - \frac{\ddot{R}R}{\dot{R}^2}.
\end{equation}

The result \maria{for both models} is presented in Fig.~\ref{fig:4} where the present-day values of the parameter are shown by black dotes. It is seen from the figure that the deceleration parameter $q(R)$ is negative for each of the galaxies at all the distances and time moments under consideration. This indicates that the outer part of the Sculptor Filament is accelerating and the dark energy dominates its dynamics since the present epoch at least. We can also see that at larger distances when the mutual gravity vanished the trajectories tend to the asymptotic value $q(R)_\infty = -1$. This local asymptotic value is the same as the cosmological one $q(t)_\infty$.

\section{Conclusions}
\label{sec:conclusion}

Dark energy was first observed with the Hubble Space Telescope at the global cosmological distances $\sim$1000 Mpc~\citep{riess1998,perlmutter1999}. The results have later been confirmed (and refined) in the space investigations with the WMAP~\citep{2013ApJS..208...19H} and Planck~\citep{Planck2016} missions. Dark energy  manifests itself only by its anti-gravity. This new force of nature was predicted theoretically by Einstein more than a hundred years ago. In his General Relativity equations of 1917, anti-gravity is represented by the cosmological constant $\Lambda$. A possible existence in nature of Einstein's anti-gravity  was taken into account in Friedmann's global cosmological theory; the cosmological constant is an empirical parameter that should be measured in astronomical observations. General Relativity indicates that dark energy exists not only in the space of the global Universe, where it was first observed, but actually everywhere in the physical space, including the local distances of 1--30 Mpc where the  distribution of galaxies is, generally, non-uniform. Since Friedmann's theory does not work there, a theory that would take into account the really observed distribution of galaxies in different space volumes is needed. 

The building  blocks of the Cosmic Web are 3D groups and clusters of galaxies, 2D walls like the Local Supercluster and 1D filaments. These systems are observed on both global and local spatial scales.  All of them are embedded in the omnipresent dark energy background. We studied the  nearest 3D expansion flows around the groups (the Local Group and several similar groups) and the Coma and Virgo clusters~\citep{chernin2001,chernin2008,2013JETPL..98..353C}; 2D Zel'dovich Local Pancake~\citep{chernin2015}. The HST recent accurate data on these systems are provided by Karachentsev and his co-workers~\citep{2003A&A...404...93K,2013AJ....145..101K}.  In this work the N-body model for 1D Sculptor Filament of galaxies has been constructed. \maria{We performed the calculations in two versions. The main model included the influence of the Virgo and Fornax clusters. In the other version, only the filament members were considered. We showed that the dark energy domination in the Sculptor Filament is not strongly affected by the most massive nearby clusters of galaxies. However, their gravity slightly modifies the spatial trajectories of the filament members.} Both computer models demonstrated that the motions of outer parts of the filament are driven by the dark energy  while the mutual gravity of the bodies is relatively weak. On the contrary the inner part is subjected to the mutual gravity and point-like approximation does not work for it. Therefore, the outer part of the filament as well as 2D and 3D systems reveal a common feature: in all the systems, the dark energy anti-gravity is stronger than the self gravity of the galaxies. Because of this, the objects in the Sculptor Filament expand with acceleration tending with time to the linear velocity-distance relation with $H_0$ value close to the current one.

\section{Compliance with Ethical Standards}
 The manuscript complies to the Ethical Rules of Astrophysics and Space Science. There are no any conflicts of interest.

 \acknowledgments

We are grateful to N.~V.~Emelyanov, Yu.~N.~Efremov, G.~S.~Bisnovatyi-Kogan, and A.~V.~Zasov for helpful discussions. M.V.P. acknowledges support from Russian Science Foundation grant 18-72-00159  for the Hubble diagram analysis and Lomonosov Moscow State University Program of Development "Leading Science Schools of MSU: Physics of Stars, Relativistic Compact Objects and Galaxies" for the $\Lambda$N-body simulations.

%% References
% \bibliographystyle{spr-mp-nameyear-cnd}  %% BibTeX style
 \bibliographystyle{plainnat}  %% BibTeX style
 \bibliography{Pruzhinskaya_biblio}                %% BibTeX data

\end{document}